\documentclass[nonacm,sigplan]{acmart}

\usepackage[]{hyperref}
\usepackage{url}
\usepackage{float}
\usepackage{caption}
\usepackage{multirow}
\usepackage{booktabs}
\usepackage{makecell}
\usepackage{graphicx}
\usepackage[ruled,vlined]{algorithm2e}
\usepackage{stfloats}
\usepackage{threeparttable}
\usepackage{pifont} 
\usepackage[mathscr]{euscript}
\usepackage{utfsym}
\usepackage{bbding}
\usepackage{threeparttable}
\usepackage{enumitem}

\author{Congcong Chen, Jihua Cui, Jiliang Zhang}
\authornote{Corresponding author. This manuscript was first submitted to the ACM International Conference on Architectural Support for Programming Languages and Operating Systems on October 18, 2024 (Fall Cycle)}
\affiliation{
  \institution{Hunan University}
  \city{Changsha}
  \country{China}
  \postcode{410082}
}
\email{zhangjiliang@hnu.edu.cn}

\begin{document}
\title{BETA: Automated Black-box Exploration for Timing Attacks in Processors}

\newcommand{\todo}[1]{\textcolor{blue}{TODO: #1}}
\newcommand{\modi}[1]{\textcolor{red}{#1}}
\newcommand{\attack}{{\sc Sync+Sync}}
\newcommand{\att}{\bfseries{\sc Sync+Sync}}
\newcommand\fixme[1]{\textcolor{red}{#1}}

\newcommand*\fullcirc[1][0.5ex]{\tikz\fill (0,0) circle (#1);}
\newcommand*\emptycirc[1][0.5ex]{\tikz\draw (0,0) circle (#1);} 

\begin{abstract}


Modern processor advancements have introduced security risks, particularly in the form of microarchitectural timing attacks. High-profile attacks such as Meltdown and Spectre have revealed critical flaws, compromising the entire system’s security. Recent black-box automated methods have demonstrated their advantages in identifying these vulnerabilities on various commercial processors. However, they often focus on specific attack types or incorporate numerous ineffective test cases, which severely limits the detection scope and efficiency.

In this paper, we present BETA, a novel black-box framework that harnesses fuzzing to efficiently uncover multifaceted timing vulnerabilities in processors. Our framework employs a two-pronged approach, enhancing both mutation space and exploration efficiency: 1) we introduce an innovative fuzzer that precisely constrains mutation direction for diverse instruction combinations, including opcode, data, address, and execution level; 2) we develop a coverage feedback mechanism based on our instruction classification to discard potentially trivial or redundant test cases. This mechanism significantly expands coverage across a broader spectrum of instruction types. We evaluate the performance and effectiveness of BETA on four processors from Intel and AMD, each featuring distinct microarchitectures. BETA has successfully detected all x86 processor vulnerabilities previously identified by recent black-box methods, as well as $8$ previously undiscovered timing vulnerabilities. BETA outperforms the existing state-of-the-art black-box methods, achieving at least $3$x faster detection speed.


\end{abstract}

\maketitle 
\pagestyle{plain} 

\section{Introduction}

Modern processors, which are at the core of computer systems, use advanced microarchitectural features like simultaneous multi-threading (SMT) and speculative execution to handle growing performance and functionality needs. However, this complexity in hardware design has led to a wider range of verification and protection requirements, making security challenges more severe. One significant concern is microarchitectural timing vulnerabilities, as demonstrated by high-profile attacks like Meltdown and Spectre. Furthermore, the emergence of related attacks, including Ret2spec \cite{Ret2spec}, Fallout \cite{Fallout}, RIDL \cite{RIDL}, ZombieLoad \cite{ZombieLoad}, LVI \cite{LVI}, and CacheOut \cite{CacheOut}, has raised serious data leaks from processors. Thus, finding timing vulnerabilities in processors is imperative for ensuring hardware security.


The discovery of timing vulnerabilities typically entails a manual process that is not only labor-intensive and time-consuming, but also demands a high level of expertise in microarchitectural design. To address this issue, researchers are increasingly turning to automated methods for vulnerability detection. There are three main categories: white-, gray-, and black-box approaches. The white-/gray-box enables early detection during the design phase. However, they are limited in their applicability across different platforms because they rely heavily on detailed processor knowledge \cite{lost,TEESec} or manual model creation and property proof\cite{formal,UPEC}. In contrast, black-box approaches, which analyze fabricated chips, have shown success in identifying timing vulnerabilities across various x86 processors \cite{Hide}.


However, the black-box approaches have severe limitations: 1) they rely on mutating known code templates (e.g., transient or side-channel gadgets), which is effective for identifying variants of known attacks, yet falls short in revealing new leaks \cite{Medusa,Template}; 2) they often focus on specific microarchitectural components (e.g., caches) rather than examining the entire processor\cite{Hide,AutoCAT}; 3) most black-box methods are tailored to detect only certain types of attacks (e.g., contention-based side channels), which restricts their scope and ability to discover other types of attacks, such as eviction-based ones \cite{Medusa,Template,Shotgun,ABSynthe}; 4) restricting the length of instruction sequences is essential because of the numerous possible combinations of instructions and the absence of comprehensive coverage metrics \cite{Osiris}. However, previous studies have not addressed this issue.


Our goal is to identify a broader range of vulnerabilities while minimizing detection overhead and addressing existing limitations. However, achieving these goals is non-trivial and poses several challenges: 1) the timing sources of microarchitectural vulnerabilities may vary and be triggered by different instructions, making it hard to establish universal detection constraints; 2) the instructions involved in these vulnerabilities often lack syntactical correlations, making traditional hardware fuzzing tools ineffective for black-box vulnerability discovery, as these tools rely on feedback such as code path coverage or RTL registers; 3) since microarchitectural vulnerabilities are semantic flaws in nature and lack specific detection constraints, a precise timer is required for measurement. However, in a black-box environment, the system is prone to disturbances, which requires that the time differences be sufficiently large and easily discernible for accurate measurement.


In this paper, we propose BETA, a novel detection framework, designed to efficiently uncover multifaceted timing vulnerabilities in processors, thereby addressing the aforementioned challenges. 1) We analyze the source of timing differences among various attack types, considering variations in data, addresses, opcodes, and execution levels (such as single-thread, SMT, and multi-threading across cores). While the instructions for different attacks vary, they also share certain commonalities within the same type, such as the way port contention and retirement bottlenecks influence timing. Building on this insight, we design a specialized fuzzer that selects the appropriate mutation direction and applies constraints tailored to different instruction combinations. This enables BETA to detect multifaceted timing vulnerabilities, as opposed to focusing solely on a single type \cite{ABSynthe}. 2) We categorize instructions into distinct lists using machine-readable Instruction Set Architecture (ISA) specifications \cite{uops}. Our analysis shows that instructions within the same category often lead to similar vulnerabilities. Based on this result, we devise a coverage feedback mechanism to reduce invalid or redundant test cases, significantly improving detection speed compared to the unguided random method employed by Osiris \cite{Osiris}. 3) We leverage a combination of multiple confirmation and covert transmission to accurately verify vulnerable instructions. This fully automated process minimizes discernible time differences (down to $10$ cycles) and eliminates the need for complex manual verification.

BETA performs automated testing of machine-readable instructions on commercial (black-box) processors by evaluating execution time differences before and after mutations. We tested the BETA framework on four Intel and AMD processors with different microarchitectures. After over 100 hours of testing, BETA successfully discovered totally $20$ known and new side channels. The known side channels include cache-based \cite{FLUSH+RELOAD,Flush+Flush,prefetch}, AVX \cite{Netspectre}, port and retirement contention \cite{Port,retire}, and floating-point (FP) timing channels \cite{Subnormal}, along with $4$ side channels previously discovered by Daniel et al. \cite{Osiris}. Additionally, BETA uncovered $8$ new side channels, including four data channels, i.e., \texttt{I286PROTECTED}, \texttt{FMA}, \texttt{x87} and \texttt{FBLD} , two x87 floating point unit (FPU) channels, and two \texttt{RDRAND} instruction-related channels. Finally, BETA automatically synthesizes the discovered timing vulnerabilities into covert channels and transient execution attacks and verifies the exploitability of those vulnerable channels.

Our contributions can be summarized as follows:
\begin{itemize}
    \item \textbf{Comprehensive Framework.} We present BETA, a comprehensive automated framework for black-box vulnerability discovery on x86 processors. BETA is capable of automatically testing timing behaviors caused by instruction operands, contention, and state switching to identify various timing vulnerabilities.
    \item \textbf{High-efficiency Fuzzer.} We design a fuzzer that can reduce the testing space by constraining mutation directions for different instruction combinations. Furthermore, we introduce a feedback mechanism based on the coverage of instruction classification to guide the fuzzer, achieving at least 3x faster detection speed than the latest black-box method, Osiris \cite{Osiris}.
    \item \textbf{Prototype Implementation.} We prototype BETA on Intel and AMD processors with different microarchitectures. Experimental results showcase that BETA found $12$ known and $8$ new timing vulnerabilities.
    \item \textbf{Attack Verification.} BETA automatically synthesizes the discovered timing vulnerabilities into covert channels and transient execution attacks to assess their exploitability, which significantly reduces the intricacies involved in manual verification.
\end{itemize}

The remainder of this paper is organized as follows. Section II presents background information on fuzzing, microarchitectural timing channels, and transient execution attacks. Section III analyzes the timing behavior of different attack types. Our detailed framework design is introduced in Section IV. Section V evaluates the performance and effectiveness of BETA. Section VI discusses the limitations and possible future improvements of this work. Related work and comparisons are surveyed in Section VII, and we conclude this paper in Section VIII.

\section{Background}
This section provides background knowledge on fuzzing and microarchitectural timing attacks, serving as the basis for this work.
\subsection{Fuzzing}

Fuzzing is an automated testing approach that generates random data, either automatically or semi-automatically, in accordance with predefined rules. This data is then fed into the target program to observe any anomalies. The effectiveness of fuzzing tools, or fuzzers, depends on two critical components: input data generation and coverage assessment.

Given the vast input space, fuzzers have to employ appropriate generation strategies to identify inputs that may trigger bugs while avoiding uninteresting ones. Current input generation strategies primarily include: 1) Mutation-based: This strategy generates new test data by randomly or systematically modifying an initial set of inputs, known as seeds. Common mutation operations include bit flipping, byte substitution, and the addition of special characters; 2) Grammar-based: This approach utilizes predefined grammar rules to create valid input data. By constructing a grammatical description of the input, it can produce test data that conforms to the program's expectations; 3) Model-based: This strategy involves establishing a model of the program's input, often using machine learning techniques to learn input distributions from historical data, and then generating new test data based on this learned model. Additionally, fuzzing techniques can be classified into two categories based on test coverage: Blind fuzzing and coverage-guided fuzzing. The former relies solely on randomly generated input data, without considering the program's structure or behavioral characteristics. In contrast, the latter leverages execution feedback to inform and guide the input generation process.

\subsection{Microarchitectural Timing Side- and Covert Channels}
Timing covert channels exploit variations in timing or delay to transmit information, rather than relying on communication interfaces defined by the architecture or operating system (OS) \cite{chen}. In a covert channels (referring to timing in context), there are two parties involved: the sender and the receiver. Typically, the sender is a Trojan process that can steal secret data and encode it into the microarchitectural state. The receiver acts as a Spy process, responsible for decoding and recovering the secret data according to a predefined protocol. For instance, in a Flush+Reload attack \cite{FLUSH+RELOAD}, the receiver begins by using the \texttt{CLFLUSH} instruction to flush the target cache line. The receiver then waits for the sender, who determines whether to load the target data based on the secret being transferred, where ``1'' indicates loading and ``0'' indicates not loading the data. Finally, the receiver reloads the target data and measures its access time. A short access time indicates the data has been cached, revealing that the transmitted secret is ``1''. Otherwise, the secret is inferred to be ``0''. Covert channels are often employed in conjunction with other attacks, serving as leakage channels for transient execution attacks. Side-channel attacks are similar to covert channels, but the key distinction is that the sender is not a malicious entity intentionally leaking information; instead, it is a trusted entity (called the victim) that inadvertently reveals sensitive data (such as encryption keys) during execution.

\subsection{Transient Execution Attacks}
Modern processors incorporate optimization techniques such as out-of-order execution and branch prediction to significantly enhance performance. These optimization mechanisms enable the processor to execute subsequent instructions ahead of time, even if earlier instructions encounter a prediction error. When the CPU detects such an error, it discards all speculatively executed instructions to maintain the correctness of program execution. Thus, these prematurely executed instructions, referred to as transient instructions \cite{Xiong}, are not committed at the architectural level. During the execution of transient instructions, a program can perform unauthorized computations (e.g., privilege escalation) and leave measurable side effects, such as changes in cache state, which are not rolled back. This allows attackers to utilize covert channels to recover sensitive data marked during execution. This attack method is known as transient execution attacks, with notable examples including Meltdown \cite{Meltdown}, Spectre \cite{Spectre}, ret2spec \cite{Ret2spec}, and microarchitectural data sampling (MDS) \cite{RIDL,Fallout,ZombieLoad,CacheOut}, among others.

\begin{table}[t]
\small
\renewcommand{\arraystretch}{1.1}
\caption{The taxonomy of microarchitectural vulnerabilities}
\label{tab1}
\centering
\footnotesize
\begin{threeparttable}
\begin{tabular}{|m{1.3cm}<{\centering} | m{2.8cm}<{\centering} | m{0.45cm}<{\centering} | m{0.45cm}<{\centering} | m{0.45cm}<{\centering} | m{0.7cm}<{\centering}|}
\hline
{\rule{0pt}{14pt}\multirow{2}{*}[-1.5ex]{\textbf{Type}}} & \multirow{2}{*}[-3.0ex]{\textbf{Example}} & \multicolumn{4}{c|}{\textbf{Source} \tnote{$\dagger$}} \\
\cline{3-6}

& & \rotatebox{70}{\footnotesize data}
  & \rotatebox{70}{\footnotesize addr}
  & \rotatebox{70}{\footnotesize opdode} & \rotatebox{70} {\footnotesize level}\\
\hline

\multirow{3}{*}{\textbf{\makecell[c]{Data \\ channel}}} & FP data \cite{Subnormal} & \fullcirc[0.7ex] & \emptycirc[0.7ex] & \emptycirc[0.7ex] & SC\\
\cline{2-6}
 & DIV \cite{SIGFuzz} & \fullcirc[0.7ex] & \emptycirc[0.7ex] & \emptycirc[0.7ex] & SC\\
\cline{2-6}
& Compress \cite{WhisperFuzz} & \fullcirc[0.7ex] & \emptycirc[0.7ex] & \emptycirc[0.7ex] & SC\\
\hline
\multirow{4}{*}{\textbf{\makecell[c]{Volatile \\ channel}}} & Cache bank \cite{bank} & \emptycirc[0.7ex] & \fullcirc[0.7ex] & \emptycirc[0.7ex] & SC\\
\cline{2-6}
 & Execution ports \cite{Port,SMoTherSpectre} & \emptycirc[0.7ex] & \emptycirc[0.7ex] & \fullcirc[0.7ex] & SC\\
\cline{2-6}
& Memory bus \cite{bus} & \emptycirc[0.7ex] & \fullcirc[0.7ex] & \emptycirc[0.7ex] & CC\\
\cline{2-6}
& Retirement \cite{retire} & \emptycirc[0.7ex] & \emptycirc[0.7ex] & \fullcirc[0.7ex] & SC\\
\hline
\multirow{3}{*}{\textbf{\makecell[c]{Persistent \\ channel}}} & Cache-based \cite{Flush+Flush,FLUSH+RELOAD,Prime+Probe} & \emptycirc[0.7ex] & \fullcirc[0.7ex] & \fullcirc[0.7ex] & SC/CC\\
\cline{2-6}
& AVX \cite{Netspectre} & \emptycirc[0.7ex] & \emptycirc[0.7ex] & \fullcirc[0.7ex] & SC\\
\cline{2-6}
& DRAM \cite{DRAMA} & \emptycirc[0.7ex] & \fullcirc[0.7ex] & \fullcirc[0.7ex] & SC/CC\\
\hline
\end{tabular}
\begin{tablenotes} 
\footnotesize 
\item[\tnote{$\dagger$}] The source of timing differences for various covert channels. The acronym (SC: same core; CC: cross core). Symbols (\fullcirc[0.7ex] or \emptycirc[0.7ex]) indicate whether the mutation type is required.
\end{tablenotes}
\end{threeparttable}
\end{table}

Meltdown \cite{Meltdown} utilizes out-of-order execution to allow unprivileged code to access kernel memory that should be protected. When a kernel address is dereferenced, it generates a page fault. However, before the fault is handled by the CPU, the data from the kernel address is already forwarded to subsequent transient instructions. Although the results of these transient instructions are not committed to registers, they may be marked in the microarchitectural state (such as the cache), enabling attackers to extract sensitive data through cache side-channel attacks. Spectre \cite{Spectre} exploits the branch predictor to execute instructions along unintended paths, allowing speculative access to secret data. By mis-training the branch predictor, an attacker can manipulate a vulnerable branch in a victim program. When this branch is speculatively executed, it accesses the secret through an out-of-bounds array and marks the secret value in the cache. Similar to Meltdown, the attacker can recover secret data through a cache side-channel attack. Note that transient execution attacks typically use caches to encode leaked secrets, but other side channels can also be employed \cite{SMoTherSpectre, Netspectre}.

\section{Microarchitectural Timing Behaviors}

Microarchitectural timing vulnerabilities arise from various components, such as caches, execution ports, and AVX units. A comprehensive approach to detect these vulnerabilities involves generalizing potential configurations, such as threat models and cache/memory settings, for analyzing different components. However, this approach is complex and requires significant manual efforts. To address this, BETA identifies vulnerabilities at the instruction level without without the need to analyze specific component details.

The core components of an instruction are the opcode, which defines the operation to be performed, and the operands, which provide the data for the instruction. Operands can be immediate values or references to a register or memory address. Since different microarchitectural attacks exhibit unique timing behaviors, we begin by analyzing the timing sources of existing attacks. As summarized in Table \ref{tab1}, we classify these vulnerabilities into three categories: 1) Data channels: where timing variations involve instruction operands, such as differences in calculation times or memory access times for various addresses; 2) Volatile channels: where variations occur due to contention during instruction execution or memory access at different levels; 3) Persistent channels: where state transitions in the microarchitecture are influenced by instruction execution or memory access.

\begin{figure}[t]
\setlength{\abovecaptionskip}{-0.1cm}
\setlength{\belowcaptionskip}{-0.3cm}
\centering
\includegraphics[width=\linewidth]{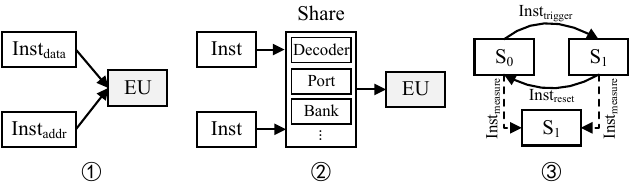}
\DeclareGraphicsExtensions.
\caption{The sources of time differences}
\label{fig1}
\end{figure}

Figure \ref{fig1} illustrates the three categories of sources of timing differences. Based on this, we can define the input space for all channels. This definition enables targeted input mutations for various instruction combinations, reducing invalid test cases and thereby facilitating more efficient leakage discovery. First, the timing differences in data channels primarily stem from the influence of instruction data and instruction addresses. Thus, the input space can be defined as follows:
\[C_d = \{ \text{data}, \text{addr} \}\]

Second, timing differences in volatile channels arise from concurrent contention for various component resources, such as ports and cache banks. This contention can be caused by address accesses, instruction opcodes, and differing levels of contention, without leaving any trace in the microarchitecture. Address access contention occurs when two addresses compete for the bandwidth of the same resource, such as a cache bank or memory bus. Instruction opcode contention arises when different instruction types contend for the same execution port. Contention levels can involve single-thread, SMT, or cross-core multi-thread scenarios. Thus, the input space can be defined as follows:
\[C_v = \{ \text{opcode}, \text{addr}, \text{level} \}\]

Persistent channels leverage the residual microarchitectural state left after instruction execution to influence subsequent instructions. The delays caused by state transitions, such as \( S_0 \rightarrow S_1 \) and \( S_1 \rightarrow S_1 \), can vary. To ensure continuous transmission, the state must be reinitialized after each execution. These channels typically involve three steps:

\begin{itemize}[leftmargin = 12pt]
\item \textbf{Reset:} This step initializes the microarchitectural state, such as evicting cache lines or powering down AVX units.
\item \textbf{Trigger:} Executed by the victim or sender, this step determines whether to execute a trigger instruction based on a secret, such as loading data into the cache or powering up AVX units. 
\item \textbf{Measurement:} Similar to the trigger instruction, this step alters the microarchitectural state. The time required for state transitions can differ, revealing whether the trigger instruction was executed based on timing differences.
\end{itemize}

These state changes can occur due to different instructions reusing resources using the same address (e.g., Flush+Reload) or contending for resources using different addresses (e.g., Prime+Probe). While these states can represent more complex abstractions, such as different cache levels, their operational mechanics are similar, and distinguishing between these two states suffices for our purposes. This contrasts with volatile channels, where cross-core and same-core contention often involve different resources and instructions. Consequently, the input space for persistent channels can be defined as: \[C_p = \{ \text{opcode}, \text{addr} \}\]

\begin{figure*}[t]
\setlength{\belowcaptionskip}{-0.2cm}
\centering
\includegraphics[width=\linewidth]{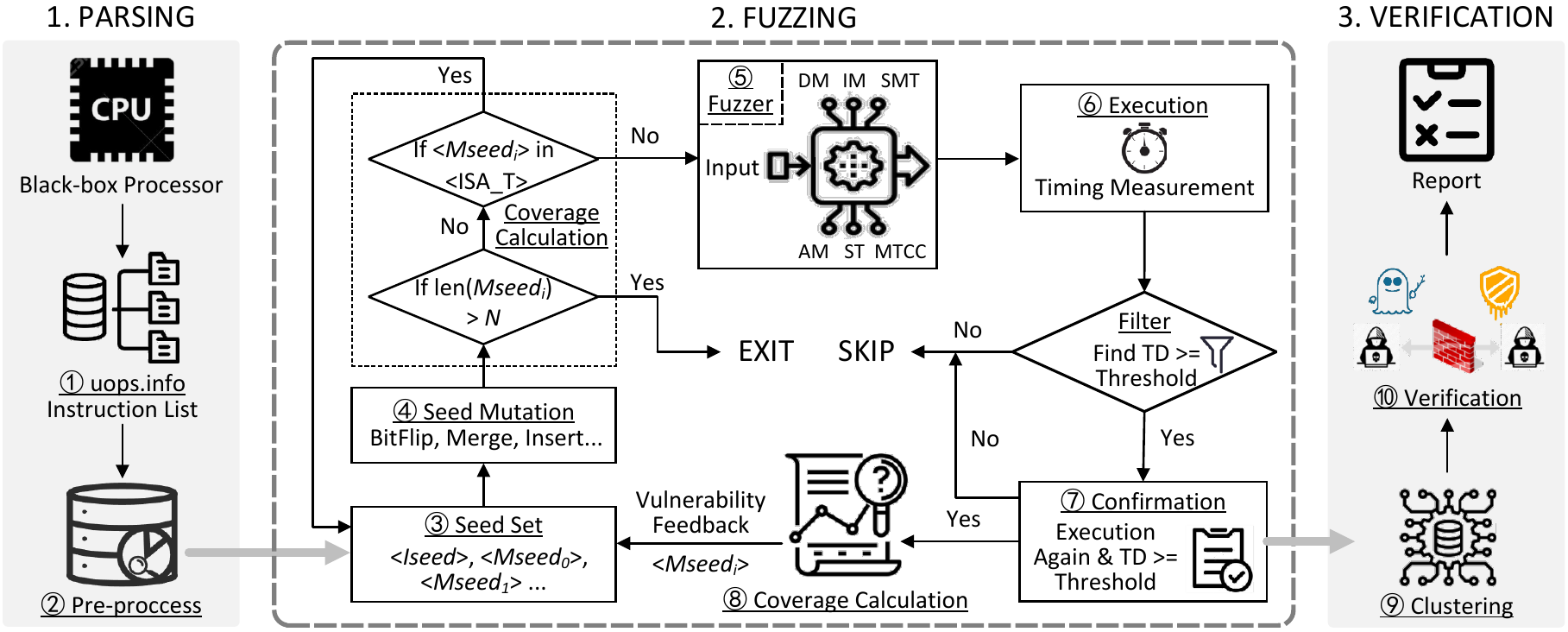}
\caption{The diagram of BETA framework. (Iseed: initial seed; Mseed: mutated seed; N: maximum instruction length; ISA\_T: tested instruction types; DM: data mutation; AM: address mutation; IM: instruction mutation; ST: single thread; SMT: simultaneous multi-threading; MTCC: multi-threading on cross cores)}
\label{fig2}
\end{figure*}

\section{BETA Design}
The BETA framework is able to cover the data, volatile, and persistent channels described above. It selects the direction of mutation for different instruction combinations depending on the timing source. As shown in Figure \ref{fig2}, BETA consists of three phases: 1) Parsing. In this initial phase, an assembly file is generated that contains all machine-readable instructions for the target processor. These instructions are generated once before the fuzzing process, reducing the overhead of generating and assembling instructions at runtime. 2) Fuzzing. This phase generates instruction combinations and purposefully mutates instructions based on different combination types. The mutated instructions are then executed, and timing differences are measured and filtered. In addition, this phase introduces a coverage feedback mechanism, which utilizes the classification of confirmed vulnerability instructions to guide the next round of seed generation. 3) Verification. In this final phase, instruction combinations that exhibited timing differences in the previous phase are validated. Clustering analysis is performed before verification to reduce overhead, because the same type of vulnerability often only needs to be verified once. Next, we will provide a comprehensive overview of the three specific phases.

\subsection{Phase 1: Instruction Parsing}
The first phase consists of the following two parts:

\textbf{Attribute Extraction.} First, we download an \texttt{XML} file from \texttt{uops.info} \cite{uops}, which contains all machine-readable instructions for various microarchitectures. This information is obtained from the configuration files of the Intel X86 Encoder Decoder (XED) library. We provide a Python script that automatically generates an instruction list from the downloaded file, including assembly code, category, extension, and ISA set (different extensions and versions of the x86 instruction set). For example, for the Intel i5-8400 processor, this list contains 14,363 variants of x86 instructions. Since BETA supports data mutation, we also extract the width and type of operands from the file. These attributes can be used to constrain the generation of instruction operands, avoiding the mutation of many invalid operands. Additionally, control flow transfer instructions (e.g., \texttt{JMP}, \texttt{RET}) are excluded from the list, as they may lead to unrecoverable states. For memory access, all registers are unified to \texttt{R8}, i.e., all memory accesses are treated as equivalent. However, the memory operands (addresses and data) pointed to by \texttt{R8} can vary. BETA also introduces a sleep pseudo-instruction \cite{Osiris}, which allows the CPU to be idle for a period. This instruction can serve as a reset step in persistent channels, such as powering down an AVX unit.

\textbf{Instruction Pre-process.} Since processors with different microarchitectures support varying instructions, BETA needs to filter out exceptional instructions automatically to adapt to the target platform. These exceptions include signals such as \texttt{SIGSEGV}, \texttt{SIGILL}, \texttt{SIGFPE}, \texttt{SIGTRAP}, and \texttt{SIGBUS}. For example, executing illegal instructions or encountering FP operation errors will trigger an exception. Since side-channel attacks typically do not cause program exceptions (excluding transient execution), the ``clean'' process is imperative. 

To simplify the process, we first initialized all registers and memory values to the same value, but it is thoughtless, as the execution of certain instructions is closely related to the specific values of their operands. For instance, the \texttt{FDIV} instruction will cause an exception if the operand is 0. The \texttt{REPE MOV} instruction causes stack smashing when the value of the \texttt{RCX} register is large. In other words, the values in memory or registers may lead to exceptions during instruction execution, but these exceptions are not caused by a lack of instruction support on the target processor. Thus, we can effectively increase the number of machine-readable instructions by reasonably initializing memory and register values. For example, for the Intel i5 8400 processor, this instruction list contains 14,363 variants of x86 instructions, and the number of clean instructions obtained under uniform initialization is 3,368. After optimizing the initialization process, the number of available instructions increased to 3,426. This indicates that a reasonable initialization strategy can significantly increase the number of executable instructions.

Additionally, we classify the instruction list based on a key observation: instructions of the same type often cause similar timing vulnerabilities. For example, most \texttt{AVX} instructions can trigger vulnerabilities related to the AVX units \cite{Netspectre}. Since there are hundreds of such instructions, repeatedly testing all of them is less efficient. Therefore, we grouped the instructions according to their ISA-set, resulting in 58 distinct instruction types. Note that a more granular classification, such as grouping by operation type (arithmetic, logic, load, store, etc.), might help uncover new vulnerabilities. However, this would significantly expand the space of instruction combinations to test. Since this is not a conceptual issue, we leave this exploration for future work.

\setlength{\textfloatsep}{5pt}
\begin{algorithm}[t]
\DontPrintSemicolon
\small
\caption{The process of seed mutation}
\label{alg1}
\KwIn{$seed\_set$: contain an initial seed $s_0$, $M$: maximum length of mutated seed; $T$: number of test instruction in a mutated seed; $R$: number of repetitions}
$ISA\_T \gets \emptyset$  // Stores already tested types \;
\While{True} {
    $tseed \gets \texttt{Random(seed\_set)}$  // Select a random seed from the seed set  \;
    \If {\texttt{randsel()}}{
        $mseed \gets BitFlip(tseed)$   // Bit flip mutation \;
    } 
    \Else { 
        $mseed \gets ConOrInsert(tseed)$  // Concatenate or insert mutation \;
    }
    // Exit if the seed reaches the maximum length\;
    \If {len($mseed$) > $M$}{
        \textbf{break}  \;
    }
    // Skip if the seed has already been tested \;
    \If {$mseed$ in $ISA\_T$}{
        \textbf{continue}  \;
    } 
    $ISA\_T \gets ISA\_T \cup \{mseed\}$ \;
    // Randomly select T instructions for execution \;
    $insts \gets RandomInstruction(mseed, T)$  \;
    $result \gets Execute(insts, R)$  // Repeat R times \; 
    // Finding vulnerabilities \;
    \If{$result > threshold$} {
        $seed\_set \gets seed\_set \cup \{mseed\}$
    }
    // The coverage of length of mutated seed \;
    \If {$Covercal$(len($mseed$)) == 100\%}{ 
        $Remove$(len($mseed$)) // Remove the length of the seed from the seed set   \;
    }
}
\end{algorithm}

\subsection{Phase 2: Fuzzing}
The second stage is fuzzing, whose goal is to acquire the instruction combination sequence with timing difference. This phase can be divided into the following six parts:

\textbf{Seed Set.} We use a seed set to store combinations of instruction types that are vulnerable. The seed set begins with an initial seed, which can be any instruction type that has been verified to contain vulnerabilities, such as x87 FP instructions. To facilitate ease of mutation and coverage calculation, we represent the seed in binary encoding. Let \( N \) be the number of instructions to be tested, each seed will need \( 6 \times N \) bits for representation due to 58 instruction types. The default maximum value of \( N \) is 50 in BETA. This is sufficient because it can cover all eviction-based channels, even some that require prime to set the cache state, such as Reload + Refresh \cite{RELOAD+REFRESH} and LRU \cite{LRU}. In addition, channels with a higher number of instructions are more difficult to exploit, and there is no evidence that they are vulnerable. For flexibility, the value of \( N \) can also be specified according to the requirement.

%
\textbf{Seed Mutation.} During each testing process, we randomly select a seed from the seed set for mutation. The mutation methods include bit flipping, concatenation, and insertion. The specific mutation process is shown in Algorithm \ref{alg1}. BETA primarily opts for bit flipping, which is chosen with a high probability of 80\%, while concatenation and insertion are selected less often, each at a probability of 10\%. This is because we prioritize coverage of channels with fewer instruction types. Bit flipping randomly chooses a Hamming distance $k$ (from 1 to seed size) to flip. To make better use of the chosen seed ($tseed$), we test all seeds that are at a Hamming distance of $k$ away from it. (the $mseed$ may contain multiple seeds in this case). Concatenation involves combining two seeds, while insertion requires placing a random 6-bit sequence at a random position within the chosen seed. Specifically, concatenation adds a random 6-bit seed to the end of the $tseed$, and insertion puts this same seed in a random spot within the $tseed$.

After obtaining a new mutated seed (not included in $ISA\_T$), BETA will test at least $T$ times within the instruction type combination. The $T$ is set to 50\% of the number of instructions within each instruction classification by default, which can be adjusted as needed. It should be noted that we do not need to complete the test $T$ times. Once a vulnerability is found during this period, BETA will exit the test of the instruction type in advance. If no timing differences are found after execution, the mutated seed will be discarded; otherwise, the seed will be added to the seed set. In addition, if a seed coverage of a particular length reaches 100\%, such as full coverage of all types of a single instruction, BETA will remove such seeds (6-bit length) from the seed set.

\textbf{Fuzzer.} Before running the fuzzer, BETA randomly selects instructions for combination based on the mutant seed. As analyzed in Section 3, different types of vulnerabilities have varying input spaces (i.e., mutation directions). We can choose the mutation direction according to the instruction combination. For a single instruction, only data channels can occur, so its mutation space is $C_{d}$. For two instructions, we conduct tests based on volatile channels with a mutation space of $C_{v}$ since they cannot form a persistent channel with the three-step sequence. To amplify the contention effect, each instruction is repeated 64 times. For three or more instructions, a persistent channel can be formed, so the mutation space is $C_{p}$. In particular, the AVX channel \cite{Netspectre} is also a three-instruction channel, as the sleep operation is considered an instruction and is added to the list of instructions.

For address mutation, BETA adds a random offset to change the address, ensuring it remains within the available address range mapped by \texttt{mmap}. This random offset has a certain probability of mapping to the same or different pages (set at 20\% and 80\%, respectively). For data mutation, BETA overwrites random normal values and special values. The special integer values include zero, maximum, and minimum values. The special FP values include subnormal, zero, infinity, NaN, maximum, and minimum values. To further increase the mutation space, we randomly initialize the normal values pointed to by memory addresses in each test. Since the trigger step in side-channel attacks is executed by the victim, a threatening attack usually aims for minimal operations by the victim. Therefore, the trigger phase typically consists of only one instruction. For opcode mutation, BETA replaces the trigger instruction with \texttt{nop} to observe its effect on subsequent measurement instructions, thus avoiding a higher false positive rate that might arise from replacing multiple instructions at once. This structured approach to fuzzing allows for comprehensive analysis of instruction interactions and potential vulnerabilities while managing the complexity of instruction mutations effectively.

\begin{figure}[t]
\centering
\includegraphics[width=0.95\linewidth]{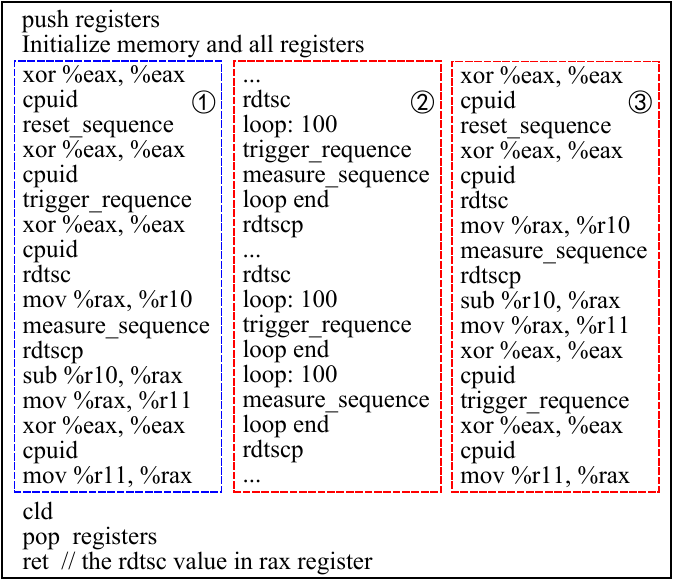}
\DeclareGraphicsExtensions.
\caption{The code page for execution. The blue box \ding{172} tests the effect of residual state between instructions. The red boxes \ding{173} and \ding{174} test the effect of contention between instructions (where \ding{173} represents port contention and \ding{174} represents out-of-order execution contention)}
\label{fig3}
\end{figure}

\textbf{Instruction Execution.} To accurately measure the execution time of instructions, it is essential to reduce external interferences, such as interrupts and dynamic voltage frequency scaling. We first set the processor to ``PERFORMANCE'' mode and disable Intel Turbo Boost. Then, we isolate two dedicated cores and bind BETA to these cores to prevent core migration. The testing code is placed on two specialized code pages, and the instruction data is placed on a specialized data page. Note that volatile channels require two data pages to avoid the effects of data contention (multiple threads accessing an object at the same time). Additionally, in the persistent channel, we point the memory-accessed register \texttt{R8} to a different location than the other registers to avoid false positives due to data dependence.

As shown in Figure \ref{fig3}, the instruction combination is divided into three parts: reset sequence, trigger sequence, and measurement sequence. For data channels, the reset and trigger sequences are empty. For volatile channels, only the reset sequence is empty. BETA tests volatile channels using both single-threading and multi-threading. Single-threading can test channels caused by port contention \cite{smt} and out-of-order execution contention \cite{SIA}, where memory access instructions do not need to be considered. Multi-threading tests channels caused by contention on other components, such as cache bank \cite{bank} and retirement units \cite{retire}. In addition, memory serialization instructions (e.g., \texttt{cpuid} and \texttt{mfence}) are added between each sequence to guarantee precise timing. 

\begin{figure}[t]
\centering
\includegraphics[width=0.95\linewidth]{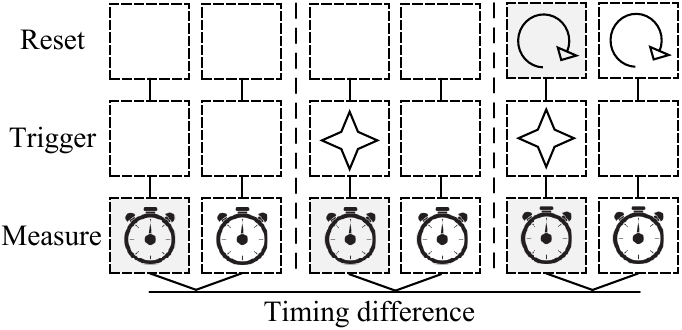}
\DeclareGraphicsExtensions.
\caption{Comparison of two code execution paths (gray areas indicate potential data or address mutations)}
\label{fig4}
\end{figure}

Next, this code is executed twice, as shown in Figure \ref{fig4}. The first execution uses original data/original addresses with a trigger, while the second one uses mutated data/mutated addresses without a trigger. BETA compares the timing differences between the two executions and records them for filtering in the next phase. To further minimize system interference, the code can be repeated $R$ times (default is 10) to take the median time. A larger $R$ value increases measurement accuracy but also prolongs the test time, allowing for a trade-off between accuracy and efficiency.

\textbf{Filter.} This process is responsible for screening out instruction combinations with large time differences, and recording the corresponding mutation data and addresses. The time difference threshold can either match the value used in the confirmation process or be set lower to capture more potential vulnerabilities. In this way, BETA can perform an initial rough screening followed by a more precise filtering during the confirmation phase.

\textbf{Confirmation.} We incorporate a confirmation procedure to reevaluate the filtered instruction combinations, further validating the suspicious instructions. This confirmation process enables BETA to detect vulnerabilities with small time differences with high accuracy. For single and double instructions, the threshold is set to $10$ cycles, while for multiple instructions, it is $20$ cycles. Smaller differences are challenging to distinguish and exploit in practice due to fluctuations caused by system or internal processor interference.

\textbf{Coverage Feedback.} Since the same instruction types often lead to similar timing vulnerabilities, we implement a coverage feedback mechanism after categorizing instructions. If a vulnerability is identified in a combination of instruction types, there is no need to retest that type. During the testing of each seed, BETA records the combinations of instruction types that have already been tested and adds them to $ISA\_T$. When a vulnerability is confirmed for a specific instruction type, the corresponding seed is incorporated into the seed set. After the test is completed, the instruction type coverage is calculated. If the coverage reaches 100\%, it indicates that all instruction types of that length have been tested and can be removed from the seed set. Therefore, combinations of instructions that have already been tested will not be repeated in subsequent tests. This coverage feedback mechanism significantly filters out redundant or duplicate test instructions, thereby enhancing inspection efficiency.

The BETA framework repeats these above steps until there is a timeout (the pre-set time), or the mutated seed $mseed$ exceeds the maximum length $M$.

\subsection{Phase 3: Verification}

The third phase can be performed either online or offline. When no address mutations are involved, it can be carried out entirely offline. Once the verification phase is completed, BETA will report on the combination of instructions with timing vulnerabilities and exploitable attacks.

\textbf{Clustering.} In this step, BETA groups the inputs forwarded from the verification phase into clusters representing different side channels. Clustering is based on various attributes of the instruction sequences, such as instruction extensions, memory behavior, and instruction categories. Our tests show that timing differences are often a critical clustering attribute, as similar side channels tend to exhibit comparable timing variations across corresponding instructions. Furthermore, if high-performance counters are available \cite{MCSCA}, we can also combine them to achieve dynamic monitoring, as channels that cause similar microarchitectural effects can often be grouped together.

\textbf{Attack Verification.} This phase validates covert channels and transient execution attacks. The results of covert channel verification can serve as an additional confirmation step. For instance, we consider channels with an accuracy rate below 70\% to be unusable, which may result from system interference or the inherent instability of the discovered channel. After covert channel validation, BETA screens out vulnerable instructions more accurately. In transient execution attack verification, BETA assesses whether covert channels can be used for information transmission during transient execution by placing trigger sequences during speculative execution. We use Retpoline \cite{Retpoline} to trigger speculative execution, as it avoids the need for the procedure of branch mistraining.

\section{Evaluation}
In this section, we evaluate the performance and effectiveness of BETA. We implemented a prototype on four different processor platforms, including three desktops with Intel i5-8400 (Coffee Lake), Intel i7-4790 (Haswell), and Intel Xeon 8444H (Sapphire Rapids) processors, as well as a laptop with an AMD Ryzen 5 5600U (Zen 3) processor. All machines run either Ubuntu 20.04 or Ubuntu 22.04 operating systems.

 

\subsection{Performance}
We first evaluated the performance of BETA, focusing on key factors such as the maximum number of instructions ($M$), the number of instructions per seed test ($T$), the number of repetitions ($R$), and the instruction coverage. For comparison, we followed the approach used in Osiris \cite{Osiris} by setting $M$ to $3$ and $R$ to $10$. When coverage reached $100\%$ and $T$ was $50\%$, BETA completed the test within $22$ hours. This resulted in a performance improvement of approximately $7.8$ times compared to Osiris on our platform. Even with $T$ at $100\%$, the performance improvement of BETA is still approximately $3$ times due to the coverage feedback of instruction type. Notably, BETA operates with a larger list of instructions (after cleaning is 3,426) than Osiris (3,377). Additionally, coverage of multiple instructions can be further enhanced by using an initial seed of a specified length, constraining instruction types (e.g., load and store instructions that impact cache state), or extending the duration of the fuzzing process.

\subsection{Finding Known Channels}
To demonstrate the effectiveness of BETA, we tested its ability to detect timing channels. After running BETA for nearly $three$ days, we automatically discovered $12$ known channels and $8$ previously unknown channels. Below, we will briefly introduce these side channels.

\noindent\textbf{Cache-based Channel.} Several cache-based channels have been proposed, including Flush+Reload \cite{FLUSH+RELOAD}, Flush+Flush \cite{Flush+Flush}, MOVNT-based \cite{Osiris}, and Prefetch-based \cite{prefetch} channels. These channels exploit the timing differences between cache hits and cache misses to leak sensitive information. Since these channels originate from different instruction types and exhibit sufficient timing differences, BETA successfully identified all of them.

\noindent\textbf{AVX Channel.} This channel exploits the power-saving properties of the AVX unit. After a period of inactivity, the AVX unit powers down, leading to increased execution time for subsequent \texttt{AVX} instructions. BETA successfully identified the channel using both \texttt{AVX} and \texttt{AVX2} instructions.

\noindent\textbf{FP Timing Channel.} BETA discovered the data channel introduced by the \texttt{x87} instruction when handling subnormal data, which was first proposed by Andrysco et al \cite{Subnormal}.

\noindent\textbf{Port Contention Channel.} BETA identified multiple port contention channels, including those involving commonly used instructions like \texttt{crc32} and \texttt{popcnt}, as well as channels caused by \texttt{vpermd} and \texttt{vpbroadcastd} instructions.

\noindent\textbf{Retirement Contention Channel.} In this channel, an attacker can infer instructions executed on the other logical core by monitoring the availability of the retirement unit on its own core. BETA found contention among different processes for the retirement unit.

\noindent\textbf{Four New Channels in Orisis.} BETA successfully identified four new channels reported in Orisis \cite{Osiris}.

\begin{table*}[ht]
\centering
\small
\caption{Examples of the new channels}
\label{tab2}
\begin{tabular}{ m{3.5cm}<{\centering}  m{3.7cm}<{\centering}  m{3.4cm}<{\centering}  m{3.7cm}<{\centering}  m{1.6cm}<{\centering}}
\toprule
\textbf{Name} & \textbf{Measurement} & \textbf{Reset / Original data} & \textbf{Trigger / Mutation data} & \textbf{Diff. (cycle)} \\
\midrule
\textbf{I286PROTECTED} & \texttt{LAR ECX, [R8]} & 1 $\sim$ MAX-1 & 0 & -69 \\
\textbf{AVX or FMA} & \texttt{VMULPS XMM1, XMM2, [R8]} & 1 $\sim$ MAX-1 & SUB & -101 \\
 \textbf{x87} & \texttt{FADD dword ptr [R8]} & 1 $\sim$ MAX-1 & MAX & 10 \\
 \textbf{FBLD} & \texttt{FBLD tbyte ptr [R8]} & 1 $\sim$ MAX-1 & 0 & -20 \\
\textbf{MMX-x87-x87} & \texttt{FADDP ST(1), ST(0)} & \texttt{PMULUDQ MM1, [R8]}
& \texttt{FADDP ST(1), ST(0)} & -27 \\
\textbf{x87-x87-x87} & \texttt{FISTTP dword ptr [R8]} & \texttt{FBLD tbyte ptr [R8]} & \texttt{FISTTP dword ptr [R8]} & -252 \\
\textbf{RDRAND-AVX-AVX} & \texttt{VADDPD YMM1, YMM2, [R8]} & \texttt{RDRAND CX} & \texttt{VADDPD YMM1, YMM2, [R8]} & 135 \\
\textbf{RDRAND-VERW-VERW} & \texttt{VERW word ptr [R8]} & \texttt{RDRAND CX} & \texttt{VERW word ptr [R8]} & 100 \\
\bottomrule
\end{tabular}
\end{table*}

\begin{figure*}[t]
\setlength{\belowcaptionskip}{-0.2cm}
\centering
\includegraphics[width=\linewidth]{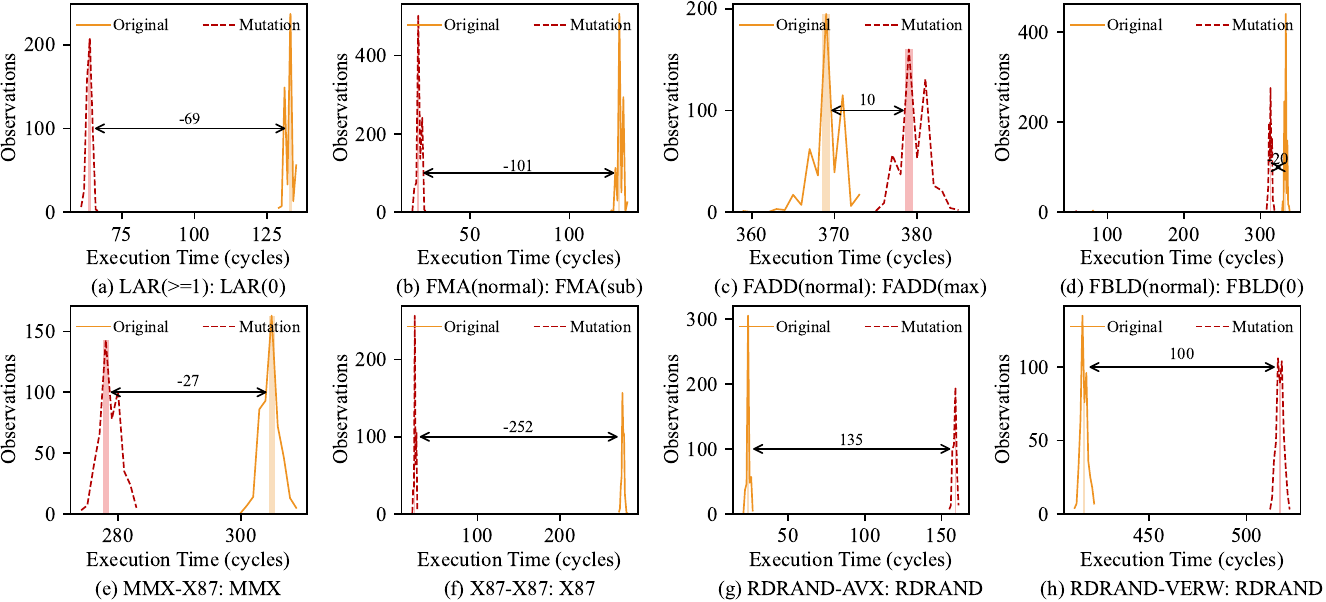}
\caption{The time difference of new channels.}
\label{fig5}
\end{figure*}

\subsection{Finding New Channels}
BETA discovered eight new channels. Table \ref{tab2} shows examples of these channels, where all the timing differences are measured on an Intel i5-8400 processor.

\noindent\textbf{\texttt{I286PROTECTED} with 0.}
Intel 80286 and later x86 processors introduced protected mode. In this mode, the processor uses several new features and instructions to support more advanced memory management and multitasking, such as \texttt{I286PROTECTED}. BETA found that for \texttt{I286PROTECTED} instructions like \texttt{LAR}, \texttt{LSL}, and \texttt{VERW}, the median difference between operands equal to and greater than 0 is $69$ cycles, as shown in Figure \ref{fig5}a. In our covert channel verification (1024 bits per transfer, repeated 10 times), this channel has an accuracy rate of $99\%$.

\noindent\textbf{\texttt{AVX} or \texttt{FMA} with SUB.}
BETA found a median difference of $101$ cycles when the operand of a \texttt{FMA} or \texttt{AVX} instruction is subnormal compared to other values, as shown in Figure \ref{fig5}b. This difference is much larger than the channels found by Andrysco et al. \cite{Subnormal}, so attackers can easily distinguish between different data. The transmission accuracy of this channel is $99.5\%$.

\noindent\textbf{\texttt{x87} with -1 or MAX.}
BETA found that when the operand for an \texttt{x87} instruction is -1 or the maximum value, a median timing difference of $10$ cycles is observed compared to other values, as shown in Figure \ref{fig5}c. This behavior is seen in instructions such as \texttt{FADD}, \texttt{FDIV}, and \texttt{FMUL}. The transmission accuracy of this channel is $99.9\%$.

\noindent\textbf{\texttt{FBLD} with 0.}
The \texttt{FBLD} instruction converts the source operand (integer) into double precision FP format and pushes the value into the FPU stack. Figure \ref{fig5}d shows that the median difference between zero and non-zero operands for this instruction is $20$ cycles. Using this channel, the transmission accuracy is $99.8\%$.

\noindent\textbf{\texttt{MMX-x87-x87} Sequence.}
Due to the shared register stack between \texttt{MMX} and \texttt{x87} instructions, switching between these two instruction types must be carefully managed to avoid inconsistent results. After executing an \texttt{MMX} instruction (except for the \texttt{EMMS}), the processor clears the state of the x87-FPU. Based on this behavior, BETA has identified a new channel: \texttt{MMX-x87-x87}. In this sequence, the \texttt{MMX} instruction first resets the x87-FPU state. If an \texttt{x87} instruction is executed during the trigger phase, the execution time is longer due to the need for state switching. Otherwise, the execution time is shorter as no state switching occurs. Figure \ref{fig5}e visualizes that the timing difference for this channel is approximately $27$ cycles, with an accuracy of $99.9\%$.


\noindent\textbf{\texttt{x87-x87-x87} Sequence.} 
The x87 register stack is an important part of the x86 architecture for FP operations. It contains eight FP registers (ST(0) to ST(7)), arranged in a hardware stack, supporting a variety of FP operations. For example, \texttt{FLD} pushes FP values into the stack, while \texttt{FSTP} pops the value at the top of the FP stack and stores it in the specified memory address. BETA discovered that instruction timing could be influenced by the presence or absence of values in the action stack, forming a sequence of such channels labeled \texttt{x87-x87-x87}. In this sequence, the reset phase pushes memory values into the stack using \texttt{FBLD} instructions. If the trigger phase uses the \texttt{FSTP} instruction to pop the values from the stack, the subsequent measurement phase's \texttt{FSTP} instruction executes more quickly. In our measurements, this channel shows a time difference of approximately $252$ cycles (Figure \ref{fig5}f) and a transmission accuracy of $99.4\%$.

\noindent\textbf{\texttt{RDRAND-AVX-AVX} or \texttt{RDRAND-VERW-VERW} Sequence.} The \texttt{RD-\allowbreak\ RAND} instruction retrieves a random value from the random number generator. This will idle the execution unit for a while, making it an effective reset instruction. BETA has identified two new classes of instructions that can be combined with \texttt{RDRAND}. The first class involves \texttt{AVX}, where \texttt{RDRAND} is used to quickly power down the AVX unit, significantly faster than sleep-based AVX channels. Using this channel, a time difference of $135$ cycles is observed (Figure \ref{fig5}g) with an accuracy of $99.9\%$. The second class involves \texttt{VERW}, where the \texttt{RDRAND-VERW-VERW} sequence shows a timing difference of $100$ cycles (Figure \ref{fig5}h) with transmission accuracy of $99.9\%$.

\subsection{Exploitability of Detected Vulnerabilities}
In this section, we verify that the suspicious instructions detected by BETA can be exploited for information leakage in different scenarios, as described below.

\textbf{Covert Channels.} In a timing covert channel, two processes controlled by an attacker (i.e., sender and receiver) breach the system isolation guarantee. The sender encodes the secret data by altering the microarchitecture state, while the receiver deduces the secret value based on observed timing differences. As discussed in Section 5.3, the identified instructions can be exploited as a covert channel. Moreover, the accuracy of these channels is relatively high. In particular, the \texttt{RDRAND-AVX-AVX} channel achieved a transmission rate of $591.7$ kbps on our system, which is about $164$ times faster than the original rate (3.6 kbps) of \texttt{Sleep-AVX-AVX} channel \cite{Netspectre}. Covert channels are often used to transmit secret data across sandbox environments, which is a common scenario in multi-tenant cloud environments.

\textbf{Transient Execution Attacks.} BETA automatically synthesizes the discovered vulnerabilities into transient execution attacks, which require the ability to either leave microarchitectural traces or compete for shared microarchitectural resources during transient execution. However, not all covert channels can be combined with transient execution attacks. Previous research has demonstrated that certain covert channels can be integrated with Spectre, such as cache \cite{Spectre}, port contention \cite{SMoTherSpectre}, and vector instructions \cite{Netspectre}. Drawing from prior work, we employed the same gadgets as Retpoline \cite{Retpoline} in BETA to automate the verification of discovered timing vulnerabilities. Our results show that the newly identified RDRAND-based channels can indeed be leveraged in Spectre-type attacks.


\section{Discussion}

\textbf{Instruction Classification Method.} BETA classifies instructions based on the ISA-set, which serves as a basic classification approach. A more granular classification could consider different types of operations within each instruction category, such as load and store operations, memory and register operations. While this refined classification may help uncover new types of vulnerabilities, it also significantly expands the testing space.


\textbf{Instruction Type Coverage.} BETA reduces the test space by classifying instructions. However, covering all possible combinations of instructions remains difficult, especially when there are a large number of instructions. The number of instructions tested in each combination, referred to as $T$, is a key factor. Determining the optimal size of $T$ requires further investigation. Nevertheless, even when T is set to its maximum value, BETA can still significantly improve performance by identifying vulnerabilities and exiting early.

\textbf{Eviction-based Channels.} While we have yet to uncover eviction-based side channels due to time constraints and the unpredictable nature of address mutations, this is not a shortcoming of our framework. BETA is capable of investigating these channels by adjusting the seed length and constraining instruction types (e.g., limiting to cache-related instructions such as load and store).


In summary, BETA is a flexible heuristic method designed to efficiently identify a variety of timing channels. Both hardware vendors and software developers can tailor the framework to suit their specific needs, enabling targeted explorations into the types of channels that are of greatest interest to them.

\section{Related Work}
Existing automated vulnerability discovery works can be divided into white- / gray-box and black-box approaches, as shown in Table \ref{tab3}.

\textbf{White- / Gray-box Approach.} Existing state-of-the-art techniques for white- and grey-box methods primarily rely on formal approaches and fuzzing techniques. However, these methods still exhibit critical shortcomings. Formal approaches, such as UPEC \cite{UPEC}, employ SAT-based bounded model checking to detect side channels in RTL designs. Although effective in exploring the design space and providing strict guarantees for design security, formal methods are hindered by the state explosion problem, making them computationally challenging for complex modern hardware. Additionally, formal methods require a thorough understanding of security specifications and the manual definition of attributes, which is an error-prone process. Grey-box fuzzers like SIGFuzz \cite{SIGFuzz} detect timing vulnerabilities in RTL by generating instruction combinations to identify cyclically precise microarchitectural timing side channels. However, the substitution of instructions can introduce architectural differences, such as variations in general-purpose registers, resulting in a high false-positive rate. WhisperFuzz \cite{WhisperFuzz}, a white-box fuzzer with static analysis capabilities, can detect and locate timing vulnerabilities but focuses only on timing differences caused by data-related inputs, without considering the interaction between instructions.

\begin{table}[tbp]
\footnotesize
\renewcommand{\arraystretch}{1.1}
\caption{Comparison with prior works. (TSC: Timing side channel)}
\label{tab3}
\centering
\begin{tabular}{|m{2.1cm}<{\centering} | m{0.8cm} <{\centering} | m{1.15cm} <{\centering} | m{0.9cm}<{\centering}| m{1.4cm} <{\centering}|}
\hline
\textbf{Paper} & \textbf{Black-box}  & \textbf{Coverage feedback} & \textbf{Non-manual} & \textbf{Target vulnerability}\\
\hline
\textbf{UPEC \cite{UPEC}} & \XSolidBrush  &  \XSolidBrush & \XSolidBrush & Covert channels  \\
\hline
\textbf{SIGFuzz \cite{SIGFuzz}} & \XSolidBrush & \CheckmarkBold & \CheckmarkBold & TSC \\
\hline
\textbf{WhisperFuzz \cite{WhisperFuzz}} & \XSolidBrush & \CheckmarkBold & \CheckmarkBold  & TSC\\
\hline
\textbf{CheckMate \cite{CheckMate}} & \CheckmarkBold  & \XSolidBrush & \XSolidBrush & Cache TSC \\
\hline
\textbf{ABSynthe \cite{ABSynthe}} & \CheckmarkBold  & \XSolidBrush & \CheckmarkBold & Contention-based TSC \\
\hline
\textbf{AutoCAT \cite{AutoCAT}} & \CheckmarkBold & \XSolidBrush & \CheckmarkBold  & Cache TSC \\
\hline
\textbf{Osiris \cite{Osiris}} & \CheckmarkBold  & \XSolidBrush & \CheckmarkBold & Eviction-based TSC \\
\hline
\textbf{BETA} & \CheckmarkBold  & \CheckmarkBold & \CheckmarkBold & TSC\\
\hline
\end{tabular}
\end{table}

\textbf{Black-box Approach.} Osiris \cite{Osiris} provides a baseline for the technique described in this paper. It introduces a method for automatically finding microarchitecture side channels using a generic three-step model, but relies on random instruction generation. To limit the search space, Osiris restricts the instruction sequence length to one, leaving vulnerabilities that require multiple instructions or specific operands to be triggered undetected. BETA, by contrast, trims duplicate or invalid test cases with coverage feedback and mutation-direction constraints, dramatically reducing test time. In addition, BETA supports data and address mutations in black-box processors, increasing the types of channels covered. Covert Shotgun \cite{Shotgun} and ABSynthe \cite{ABSynthe} automatically detect contention-based side channels that occur when instructions are executed simultaneously on an SMT core. Covert Shotgun runs a carefully selected set of instructions and measures any observable contention of one instruction against another. ABSynthe improves Covert Shotgun by automatically creating leak maps that capture interactions between different x86 instructions, and then using these maps to synthesize an attack on a given target program. AutoCAT \cite{AutoCAT} introduced a reinforcement learning-based framework for automatically generating a cache timing attack sequence for any given cache. However, its ability to detect side channels in other microarchitecture components is limited.  CheckMate \cite{CheckMate} uses a ``microarchitecturally happens-before'' graph to capture microarchitecture events and assess whether a microarchitecture is vulnerable to a given class of security vulnerabilities. In addition, it supports the automatic synthesis of exploit programs from microarchitecture and exploit pattern specifications. However, this approach relies on the matching pattern of vulnerable instructions.


In contrast, BETA addresses these limitations by offering an efficient black-box fuzzing tool that provides multifaceted coverage of vulnerabilities, including those arising from operand variations and instruction interactions within microarchitectural behaviors.

\section{Conclusion}
This paper proposes BETA, an automated black-box fuzzing framework designed to uncover microarchitectural timing vulnerabilities. BETA incorporates a novel fuzzer featuring constrained mutation direction and coverage feedback, enabling comprehensive detection of various microarchitectural timing vulnerabilities. In our evaluation, BETA uncovered 8 new timing vulnerabilities across x86 processors with four distinct microarchitectures and successfully reproduced 12 previously discovered vulnerabilities. Furthermore, BETA assesses the exploitability of these vulnerabilities by synthesizing covert channels and transient execution attacks. This approach significantly enhances the efficiency of vulnerability detection in processor microarchitectures. We aim to initiate a comprehensive and efficient approach for identifying and addressing vulnerabilities in black-box processors, thereby fostering further investigation in processor security.


\bibliographystyle{plain}
\bibliography{references}

\begin{thebibliography}{10}

\bibitem{uops}
Andreas Abel and Jan Reineke.
\newblock uops.info: Characterizing latency, throughput, and port usage of instructions on intel microarchitectures.
\newblock In {\em ASPLOS}, pages 673--686, 2019.

\bibitem{Port}
Alejandro~Cabrera Aldaya, Billy~Bob Brumley, Sohaib ul~Hassan, Cesar Pereida~García, and Nicola Tuveri.
\newblock Port contention for fun and profit.
\newblock In {\em S\&P}, pages 870--887, 2019.

\bibitem{Subnormal}
Marc Andrysco, David Kohlbrenner, Keaton Mowery, Ranjit Jhala, Sorin Lerner, and Hovav Shacham.
\newblock {On subnormal floating point and abnormal timing}.
\newblock In {\em S\&P}, pages 623--639, 2015.

\bibitem{SIA}
Mohammad Behnia, Prateek Sahu, Riccardo Paccagnella, Jiyong Yu, Zirui~Neil Zhao, Xiang Zou, Thomas Unterluggauer, Josep Torrellas, Carlos Rozas, Adam Morrison, Frank Mckeen, Fangfei Liu, Ron Gabor, Christopher~W. Fletcher, Abhishek Basak, and Alaa Alameldeen.
\newblock Speculative interference attacks: breaking invisible speculation schemes.
\newblock In {\em ASPLOS}, page 1046–1060, 2021.

\bibitem{ABSynthe}
Gras Ben, Giuffrida Cristiano, Kurth Michael, Bos Herbert, and Razavi Kaveh.
\newblock Absynthe: Automatic blackbox side-channel synthesis on commodity microarchitectures.
\newblock In {\em NDSS}, pages 1--18, 2020.

\bibitem{SMoTherSpectre}
Atri Bhattacharyya, Alexandra Sandulescu, Matthias Neugschwandtner, Alessandro Sorniotti, Babak Falsafi, Mathias Payer, and Anil Kurmus.
\newblock Smotherspectre: exploiting speculative execution through port contention.
\newblock In {\em CCS}, pages 785--800, 2019.

\bibitem{WhisperFuzz}
Pallavi Borkar, Chen Chen, Mohamadreza Rostami, Nikhilesh Singh, Rahul Kande, Ahmad-Reza Sadeghi, Chester Rebeiro, and Jeyavijayan Rajendran.
\newblock {WhisperFuzz}: {White-Box} fuzzing for detecting and locating timing vulnerabilities in processors.
\newblock In {\em USENIX Security}, pages 5377--5394, 2024.

\bibitem{RELOAD+REFRESH}
Samira Briongos, Pedro Malag{\'o}n, Jos{\'e}~M Moya, and Thomas Eisenbarth.
\newblock Reload + refresh: Abusing cache replacement policies to perform stealthy cache attacks.
\newblock In {\em USENIX Security}, pages 1967--1984, 2020.

\bibitem{Fallout}
Claudio Canella, Daniel Genkin, Lukas Giner, Daniel Gruss, Moritz Lipp, Marina Minkin, Daniel Moghimi, Frank Piessens, Michael Schwarz, Berk Sunar, et~al.
\newblock {Fallout: Leaking data on meltdown-resistant CPUs}.
\newblock In {\em CCS}, pages 769--784, 2019.

\bibitem{formal}
Mohammad~Rahmani Fadiheh, Johannes Müller, Raik Brinkmann, Subhasish Mitra, Dominik Stoffel, and Wolfgang Kunz.
\newblock A formal approach for detecting vulnerabilities to transient execution attacks in out-of-order processors.
\newblock In {\em DAC}, pages 1--6, 2020.

\bibitem{UPEC}
Mohammad~Rahmani Fadiheh, Alex Wezel, Johannes Müller, Jörg Bormann, Sayak Ray, Jason~M. Fung, Subhasish Mitra, Dominik Stoffel, and Wolfgang Kunz.
\newblock An exhaustive approach to detecting transient execution side channels in rtl designs of processors.
\newblock {\em IEEE TC}, 72(1):222--235, 2023.

\bibitem{Shotgun}
Anders Fogh.
\newblock Covert shotgun.
\newblock \url{https://cyber.wtf/2016/09/27/covert-shotgun/}, 2016.
\newblock Accessed: Auguest, 2024.

\bibitem{TEESec}
Moein Ghaniyoun, Kristin Barber, Yuan Xiao, Yinqian Zhang, and Radu Teodorescu.
\newblock Teesec: Pre-silicon vulnerability discovery for trusted execution environments.
\newblock In {\em ISCA}, pages 1--15, 2023.

\bibitem{prefetch}
Daniel Gruss, Cl\'{e}mentine Maurice, Anders Fogh, Moritz Lipp, and Stefan Mangard.
\newblock Prefetch side-channel attacks: Bypassing smap and kernel aslr.
\newblock In {\em CCS}, page 368–379, 2016.

\bibitem{Flush+Flush}
Daniel Gruss, Cl{\'e}mentine Maurice, Klaus Wagner, and Stefan" Mangard.
\newblock Flush+flush: A fast and stealthy cache attack.
\newblock In {\em DIMVA}, pages 279--299, 2016.

\bibitem{Template}
Ahmad Ibrahim, Hamed Nemati, Till Schl\"{u}ter, Nils~Ole Tippenhauer, and Christian Rossow.
\newblock Microarchitectural leakage templates and their application to cache-based side channels.
\newblock In {\em CCS}, page 1489–1503, 2022.

\bibitem{bank}
Zhen~Hang Jiang and Yunsi Fei.
\newblock {A novel cache bank timing attack}.
\newblock In {\em ICCAD}, pages 139--146. IEEE/ACM, 2017.

\bibitem{Retpoline}
Mohd Fadzil~Abdul Kadir, Jin~Kee Wong, Fauziah Ab~Wahab, Ahmad Faisal Amri~Abidin Bharun, Mohamad~Afendee Mohamed, and Aznida~Hayati Zakaria.
\newblock {Retpoline Technique for Mitigating Spectre Attack}.
\newblock In {\em ICEEE}, pages 96--101, Bandung, Indonesia, 2019.

\bibitem{retire}
Xu~Ke, Tang Ming, Wang Quancheng, and Wang Han.
\newblock Exploitation of security vulnerability on retirement.
\newblock In {\em HPCA}, 2024.
\newblock Accepted.

\bibitem{Spectre}
Paul Kocher, Jann Horn, Anders Fogh, Daniel Genkin, Daniel Gruss, Werner Haas, Mike Hamburg, Moritz Lipp, Stefan Mangard, Thomas Prescher, Michael Schwarz, and Yuval Yarom.
\newblock {Spectre Attacks: Exploiting Speculative Execution}.
\newblock In {\em S\&P}, pages 1--19, 2019.

\bibitem{Meltdown}
Moritz Lipp, Michael Schwarz, Daniel Gruss, Thomas Prescher, Werner Haas, Anders Fogh, Jann Horn, Stefan Mangard, Paul Kocher, Daniel Genkin, Yuval Yarom, and Mike Hamburg.
\newblock Meltdown: Reading kernel memory from user space.
\newblock In {\em USENIX Security}, pages 973--990, 2018.

\bibitem{Prime+Probe}
Fangfei Liu, Yuval Yarom, Qian Ge, Gernot Heiser, and Ruby~B. Lee.
\newblock Last-level cache side-channel attacks are practical.
\newblock In {\em S\&P}, pages 605--622. IEEE, 2015.

\bibitem{AutoCAT}
Mulong Luo, Wenjie Xiong, Geunbae Lee, Yueying Li, Xiaomeng Yang, Amy Zhang, Yuandong Tian, Hsien-Hsin~S. Lee, and G.~Edward Suh.
\newblock Autocat: Reinforcement learning for automated exploration of cache-timing attacks.
\newblock In {\em HPCA}, pages 317--332, 2023.

\bibitem{Ret2spec}
Giorgi Maisuradze and Christian Rossow.
\newblock {Ret2spec: Speculative Execution Using Return Stack Buffers}.
\newblock In {\em CCS}, page 2109–2122, 2018.

\bibitem{Medusa}
Daniel Moghimi, Moritz Lipp, Berk Sunar, and Michael Schwarz.
\newblock Medusa: Microarchitectural data leakage via automated attack synthesis.
\newblock In {\em USENIX Security}, pages 1427--1444, 2020.

\bibitem{lost}
Rostami Mohamadreza, Zeitouni Shaza, Kande Rahul, Chen Chen, Mahmoody Pouya, Rajendran Jeyavijayan, and Sadeghi Ahmad-Reza.
\newblock Lost and found in speculation: Hybrid speculative vulnerability detection.
\newblock In {\em DAC}, 2024.
\newblock Accepted, to appear.

\bibitem{Hide}
Oleksii Oleksenko, Marco Guarnieri, Boris Köpf, and Mark Silberstein.
\newblock Hide and seek with spectres: Efficient discovery of speculative information leaks with random testing.
\newblock In {\em S\&P}, pages 1737--1752, 2023.

\bibitem{DRAMA}
Peter Pessl, Daniel Gruss, Cl{\'e}mentine Maurice, Michael Schwarz, and Stefan Mangard.
\newblock {DRAMA: Exploiting DRAM Addressing for Cross-CPU Attacks}.
\newblock In {\em USENIX security}, pages 565--581, 2016.

\bibitem{SIGFuzz}
Chathura Rajapaksha, Leila Delshadtehrani, Manuel Egele, and Ajay Joshi.
\newblock Sigfuzz: A framework for discovering microarchitectural timing side channels.
\newblock In {\em DATE}, pages 1--6, 2023.

\bibitem{smt}
Thomas Rokicki, Cl{\'e}mentine Maurice, and Michael Schwarz.
\newblock Cpu port contention without smt.
\newblock In {\em ESORICS}, pages 209--228, 2022.

\bibitem{ZombieLoad}
Michael Schwarz, Moritz Lipp, Daniel Moghimi, Jo~Van~Bulck, Julian Stecklina, Thomas Prescher, and Daniel Gruss.
\newblock {ZombieLoad: Cross-privilege-boundary data sampling}.
\newblock In {\em CCS}, pages 753--768, 2019.

\bibitem{Netspectre}
Michael Schwarz, Martin Schwarzl, Moritz Lipp, Jon Masters, and Daniel Gruss.
\newblock {Netspectre: Read arbitrary memory over network}.
\newblock In {\em ESORICS}, pages 279--299, 2019.

\bibitem{MCSCA}
Chaoqun Shen, Congcong Chen, and Jiliang Zhang.
\newblock Micro-architectural cache side-channel attacks and countermeasures.
\newblock In {\em ASP-DAC}, pages 441--448, 2021.

\bibitem{CheckMate}
Caroline Trippel, Daniel Lustig, and Margaret Martonosi.
\newblock Checkmate: Automated synthesis of hardware exploits and security litmus tests.
\newblock In {\em MICRO}, pages 947--960, 2018.

\bibitem{LVI}
Jo~Van~Bulck, Daniel Moghimi, Michael Schwarz, Moritz Lippi, Marina Minkin, Daniel Genkin, Yuval Yarom, Berk Sunar, Daniel Gruss, and Frank Piessens.
\newblock {LVI: Hijacking Transient Execution through Microarchitectural Load Value Injection}.
\newblock In {\em S\&P}, pages 54--72, 2020.

\bibitem{RIDL}
Stephan van Schaik, Alyssa Milburn, Sebastian Österlund, Pietro Frigo, Giorgi Maisuradze, Kaveh Razavi, Herbert Bos, and Cristiano Giuffrida.
\newblock {RIDL: Rogue In-Flight Data Load}.
\newblock In {\em S\&P}, pages 88--105, 2019.

\bibitem{CacheOut}
Stephan van Schaik, Marina Minkin, Andrew Kwong, Daniel Genkin, and Yuval Yarom.
\newblock {CacheOut: Leaking data on Intel CPUs via cache evictions}.
\newblock In {\em S\&P}, pages 339--354, 2021.

\bibitem{Osiris}
Daniel Weber, Ahmad Ibrahim, Hamed Nemati, Michael Schwarz, and Christian Rossow.
\newblock Osiris: Automated discovery of microarchitectural side channels.
\newblock In {\em USENIX Security}, pages 1415--1432, 2021.

\bibitem{bus}
Zhenyu Wu, Zhang Xu, and Haining Wang.
\newblock {Whispers in the Hyper-Space: High-Bandwidth and Reliable Covert Channel Attacks Inside the Cloud}.
\newblock {\em IEEE/ACM Transactions on Networking}, 23(2):603--615, 2015.

\bibitem{LRU}
Wenjie Xiong and Jakub Szefer.
\newblock Leaking information through cache lru states.
\newblock In {\em HPCA}, pages 139--152. IEEE, 2020.

\bibitem{Xiong}
Wenjie Xiong and Jakub Szefer.
\newblock Survey of transient execution attacks and their mitigations.
\newblock {\em ACM Comput. Surv.}, 54(3):1--36, 2021.

\bibitem{FLUSH+RELOAD}
Yuval Yarom and Katrina Falkner.
\newblock Flush + reload: A high resolution, low noise, l3 cache side-channel attack.
\newblock In {\em USENIX Security}, page 719–732, 2014.

\bibitem{chen}
Jiliang Zhang, Congcong Chen, Jinhua Cui, and Keqin Li.
\newblock Timing side-channel attacks and countermeasures in cpu microarchitectures.
\newblock {\em ACM Comput. Surv.}, 56(7):1--40, 2024.

\end{thebibliography}



\end{document}